\documentclass[reprint,aps,prl,amsmath,amssymb,showpacs]{revtex4-1}

\usepackage[english]{babel}
\usepackage{siunitx}
\usepackage{graphicx}
\usepackage{lmodern}
\usepackage{xcolor}
\definecolor{linkcolor}{rgb}{0,0,0.6}
\usepackage[colorlinks=true, linkcolor= linkcolor, citecolor= linkcolor, urlcolor= linkcolor]{hyperref}

\begin{document}


\title{Creeping Avalanches of Brownian Granular Suspensions}


\author{Antoine B\'{e}rut}
 \email{antoine.berut@univ-amu.fr}
\affiliation{Aix Marseille Univ, CNRS, IUSTI, Marseille, France\\}

\author{Olivier Pouliquen}
\affiliation{Aix Marseille Univ, CNRS, IUSTI, Marseille, France\\}

\author{Yo\"{e}l Forterre}
\affiliation{Aix Marseille Univ, CNRS, IUSTI, Marseille, France\\}


\date{\today}

\begin{abstract}
We study the avalanche dynamics of a pile of micron-sized silica grains in miniaturized rotating drums filled with water. Contrary to what is expected for classical granular materials, the avalanches do not stop at a finite angle of repose. Below an angle $\theta_c$, we observe a creep regime where the piles slowly flow until they become flat. In this regime, the relaxation is logarithmic in time and is slower when the ratio between the weight of the grains and the thermal agitation is increased. We propose a simple 1D model based on Kramer's escape rate to describe this flow.
\end{abstract}




\maketitle

\section{Introduction}

Macroscopic granular materials are classically considered to be athermal, which means that their mechanical properties are not influenced by thermal agitation. For example, avalanches of dense grains immersed in fluids have been studied in various configurations~\cite{CourrechduPont2003,Pailha2008,Rondon2011}. By contrast, colloidal suspensions, made of very small objetcs, are often considered in the regime where Brownian fluctuations are dominant compared to gravitational forces~\cite{Piazza2014}. For example, their sedimentation~\cite{Royall2005,Piazza2014} and structural properties~\cite{Dullens2006,Thorneywork2017} have been widely studied. However, the flowing properties in the dense regime, where both Brownian forces and contacts between grains play a role, remains largely unexplored.\\
In this article, we study the avalanche dynamics of micro-beads, that are heavy enough to sediment in their surrounding fluid and form a well-defined pile, but small enough to be sensitive to thermal agitation. This property is quantified by their gravitational P\'{e}clet number, which is the ratio of their weight and of the Brownian thermal forces~\cite{Russel1989}:
\begin{equation}
\mathrm{Pe} = \frac{m g d}{k_{\mathrm{B}} T}
\end{equation}
where $d$ is the diameter of the particles, $m$ is their mass corrected by the buoyancy ($m = 1/6 \pi d^3 \times \Delta \rho$ where $\Delta \rho$ is the difference of density between the particles and the surrounding fluid), $g$ is the intensity of the gravity ($g \approx \SI{9.81}{\meter\per\square\second}$), $T$ is the temperature of the system ($T \approx \SI{298}{\kelvin}$), and $k_{\mathrm{B}}$ is the Boltzmann constant ($k_{\mathrm{B}} \approx \SI{1.38e-23}{\joule\per\kelvin}$).\\
This approach allows us to identify a regime, that does not exist in macroscopic granular materials, where a pile of beads flows under its repose angle. We experimentally study this flow and propose a simple model to describe it. This could bring useful insights in the field of colloidal suspension rheology, by providing a pressure-imposed set-up, complementary to the volume-imposed geometry used in most studies~\cite{Mewis2012}.

\section{Experimental set-up}

\begin{figure}[ht!]
\includegraphics[width=\columnwidth]{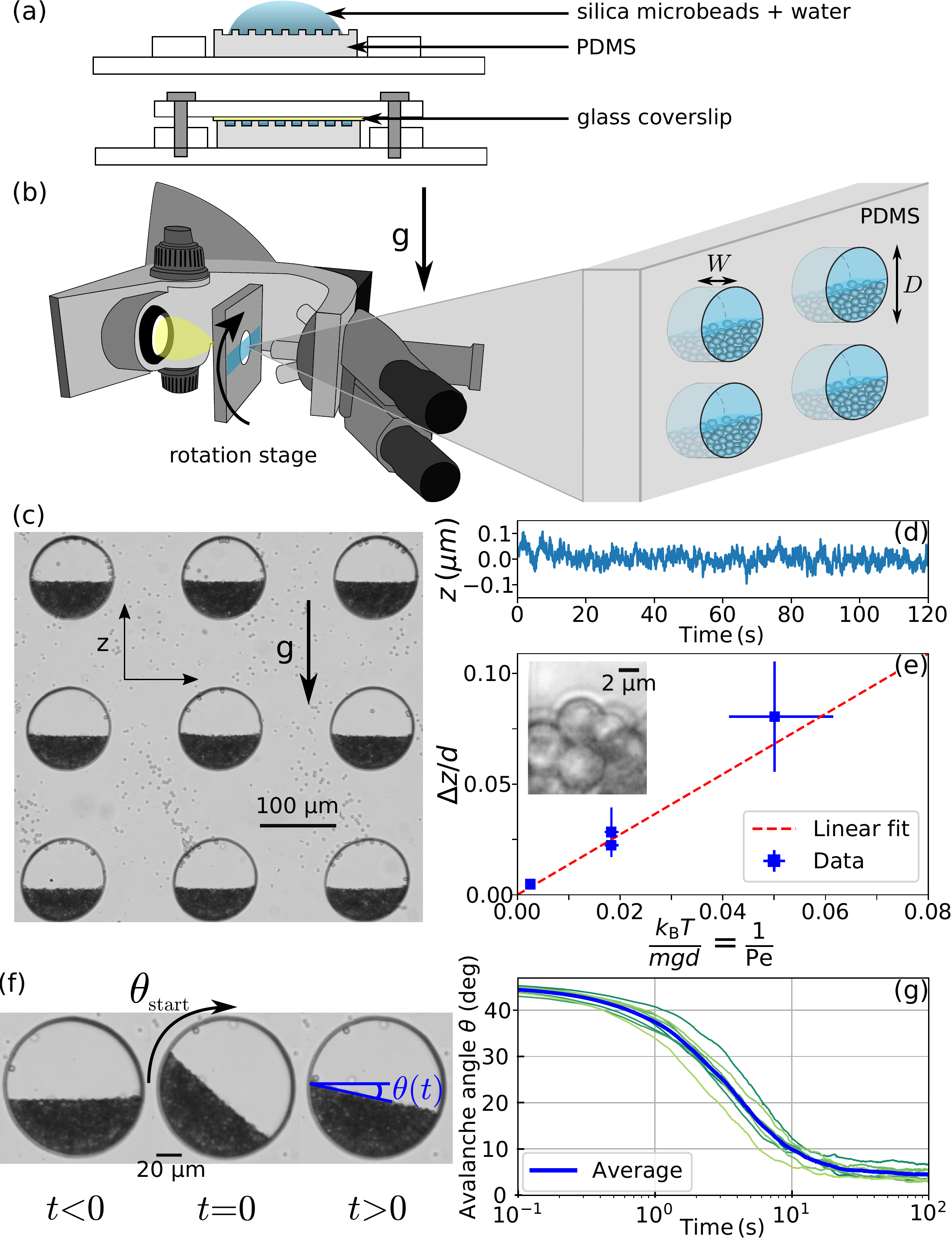}%
\caption{\textbf{(a)} Experimental sample with PDMS drum matrix held in PMMA vise (\textit{up}: before sealing, \textit{down}: after sealing). \textbf{(b)} Experimental set-up with inclined microscope. \textbf{(c)} Large view of the \SI{100}{\micro\meter} PDMS drums filled with water and \SI{4.4}{\micro\meter} silica particles. The arrow indicates the direction of gravity. \textbf{(d)} Vertical fluctuations of a \SI{4.4}{\micro\meter} silica particles at the top of the pile. \textbf{(e)} Amplitude of vertical fluctuations as a function of the P\'{e}clet number (\textit{inset}: typical image used to track single particles at the top of the pile). \textbf{(f)} Scheme of the experimental procedure: after stirring the particles are let to sediment for $\sim$\SI{5}{\minute} in the drum, at $t=0$ an initial angle $\theta_{start}$ is applied, then the angle of the piles are measured over time. \textbf{(g)} Example of pile angles temporal evolution after an initial inclination angle of \SI{45}{\degree}.\label{fig:experimental_setup}}
\end{figure}

\textit{Description.}
The experimental set-up is a polydimethylsiloxane (PDMS) matrix of micron-sized cylindrical drums (diameter $D=\SI{100}{\micro\meter}$ and width $W=\SI{50}{\micro\meter}$) filled with inert silica micro-particles dispersed in water. It is observed on a flipped microscope, so that the plane of observation is vertical and contain the gravity vector (see figure~\ref{fig:experimental_setup} (a) to (c)). We use different batches of particles with diameter size $d$ ranging from $1.55 \pm 0.05$ to $4.40 \pm 0.24$ \si{\micro\meter}, which corresponds to P\'{e}clet numbers between 6 and 400. This range of $\mathrm{Pe}$ ensures that the particles will sediment rapidly and form a well-defined pile at the bottom of the drums. However, even the bigger ones show measurable random fluctuations induced by thermal agitation (see vertical motions on figure~\ref{fig:experimental_setup} (d)).  As expected, the amplitude of the vertical fluctuations increases linearly with the P\'{e}clet, as shown on figure~\ref{fig:experimental_setup} (e). The experimental procedure is schematically shown in figure~\ref{fig:experimental_setup} (f). It consist in stirring the suspension and letting the particle sediment for $\sim$\SI{5}{\minute} to form a flat pile, then rapidly applying an initial rotation angle $\theta_{start}$, and recording the temporal evolution of the pile angle $\theta$ during the avalanche. The initial stirring ensures a reproducible initial state and avoids potential aging effects that could arise from long time contacts between grains~\cite{Gayvallet2002}. Typical avalanches trajectories with $\theta_{start} = \SI{45}{\degree}$ are shown in figure~\ref{fig:experimental_setup} (g).

\textit{Technical details.}
The matrix of cylinders is made in polydimethylsiloxane (PDMS) using standard microfluidic fabrication techniques~\cite{Duffy1998,McDonald2002,Friend2010}: a negative mold with the desired pattern was made in SU-8 photoresist at the CINaM Laboratory, then the mold was used to create a positive replica with Sylgard\textregistered 184 Silicone Elastomer Kit (standard 10:1 mixture, cured one night at \SI{60}{\celsius} in a oven). The silica particles are commercially available from Microparticles GmbH in aqueous solutions (\SI{5}{\percent}wt), and have a density $\rho \approx \SI{1850}{\kilogram\per\cubic\meter}$. The drums are filled using the following protocol: the PDMS container is first manually cleaned with isopropyl alcohol (IPA) and rinsed with pure water (Type 1 water from Purelab\textregistered Flex dispenser). Then, it is placed in the lower part of a transparent PMMA vise and washed in pure water with a ultrasonic bath for about \SI{30}{\minute}. The dispenser is rinsed again and a few droplets of pure water are put at the top of the clean PDMS surface. Then, a small volume (between 20 and \SI{40}{\micro\liter}) of silica particle solution is injected inside the water droplet. The particles are let to sediment for $\sim$\SI{2}{minutes}. A glass coverslip, previously washed with IPA, is pressed against the PDMS thanks to the upper part of the PMMA vise that is clamped with four screws (see figure~\ref{fig:experimental_setup} (a)). The coverslip is pressed carefully so that the droplet spread in the container, and no air bubble form inside. Once the container is closed, it is put vertically under the microscope to find an area where no drum is leaking (i.e. where no particle is able to escape its drum to a neighboring one). To ensure that the system is well sealed, the sample is let in vertical position for about $\sim$\SI{10}{\hour} before any measurement. This waiting time also allows for the bottom particles to stick on the drum walls, which then avoids that the particles slip on them during the avalanches. Large scale observations are made using a microscope (Leica DM 2500P) flipped horizontally with a long working distance objectives (NPLAN EPI $10 \times/0.25$ POL) allowing to watch up to 12 drums at the same time. The sample is held on a rotational stage (M-660 PILine\textregistered) with a maximal velocity of \SI{720}{\degree\per\second} and controlled by a C-867 PILine\textregistered Controller. Images and movies are taken with a Nikon D7100.

\section{Experimental observations}

\begin{figure}[ht!]
\includegraphics[width=\columnwidth]{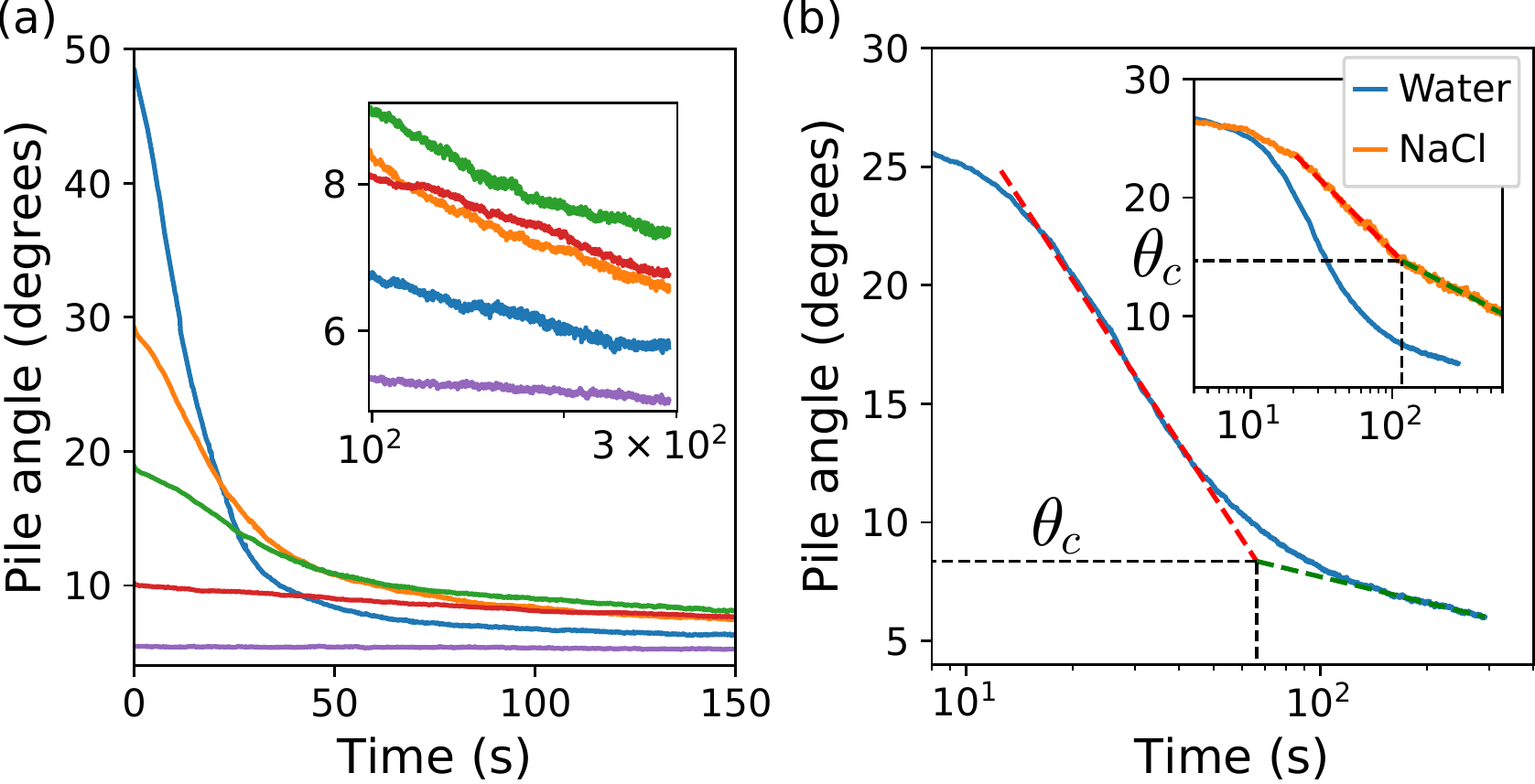}
\caption{\textbf{(a)} Avalanches for a pile of \SI{2.68}{\micro\meter} particles with different initial angles $\theta_{start}$ ranging from \SI{5}{\degree} to \SI{50}{\degree}. \textit{Inset}: Zoom on the end of the trajectories. \textbf{(b)} Avalanches for a pile \SI{2.36}{\micro\meter} particles with $\theta_{start}=\SI{30}{\degree}$ in pure water. Two regimes can be defined: the fast avalanche regime (red dashed line) and the slow creep regime (green dashed line). The threshold angle $\theta_c$ separating the two regimes is about \SI{8}{\degree}. \textit{Inset}: Comparison with the same particles in salted water (\SI{1e-2}{\mole\per\liter}). $\theta_c$ is about \SI{15}{\degree}.\label{fig:avalanches_shorttimes}}
\end{figure}

First, we verified that the piles at rest show no compaction effect. The position of the interface between a pile of \SI{2.06}{\micro\meter} particles and the fluid was tracked for \SI{2}{\hour} after stirring the drums: sedimentation was observed during $\sim\SI{3}{\min}$ and no evolution was measured afterwards.\\
Then, we considered the effect of the initial avalanches angle $\theta_{start}$ on the flowing dynamics. Avalanches for a pile of \SI{2.68}{\micro\meter} particles with $\theta_{start}$ ranging from \SI{5}{\degree} to \SI{50}{\degree} are shown in figure~\ref{fig:avalanches_shorttimes} (a). As shown in the inset, the pile seems to flow even at very small angles. In particular, a flow is still observed when $\theta_{start}=\SI{5}{\degree}$, \textit{i.e.} when the initial angle is well below the typical angle of repose for macroscopic granular materials (typically between 20 to \SI{30}{\degree})~\cite{Andreotti2013}, and even below the angle found in numerical simulations for frictionless nearly-rigid beads ($5.76 \pm 0.22\si{\degree}$)~\cite{Peyneau2008}. However, the flow at small angle seems to be much slower than the avalanche observed for high initial angles. This effect is very reproducible and was observed for all the particle sizes considered in this study. It also seems rather insensitive to aging, as \SI{48}{\hour} old samples exhibit nearly the same avalanche dynamics as the one observed at their first use.\\
By considering a semilogarithmic plot (see figure~\ref{fig:avalanches_shorttimes} (b)), one can define two regimes separated by a threshold angle $\theta_c$ : a fast avalanche regime for $\theta>\theta_c$, and a slow creep regime for $\theta<\theta_c$. For our silica particles, $\theta_c \approx \SI{8}{\degree}$, and does not seem to vary with the size of the particles. We interpret this angle as the repose angle that the granular material would have if the particles were not subjected to thermal agitation. The value of \SI{8}{\degree} is consistent with what is expected for silica particles that are nearly frictionless due to the repulsive forces between their negatively charged surfaces~\cite{Israelachvili2011}. This conclusion is also supported by the fact that $\theta_c$ is increased to $\sim$\SI{15}{\degree} if the particles are immersed in a solution of NaCl at \SI{1e-2}{\mole\per\liter} instead of pure water (see inset of figure~\ref{fig:avalanches_shorttimes} (b)). Indeed, the ions in suspension screen the repulsive forces between the particles, hence enhancing their friction, and increasing the value of the angle of repose~\cite{Clavaud2017}.

\begin{figure}[ht!]
\includegraphics[width=\columnwidth]{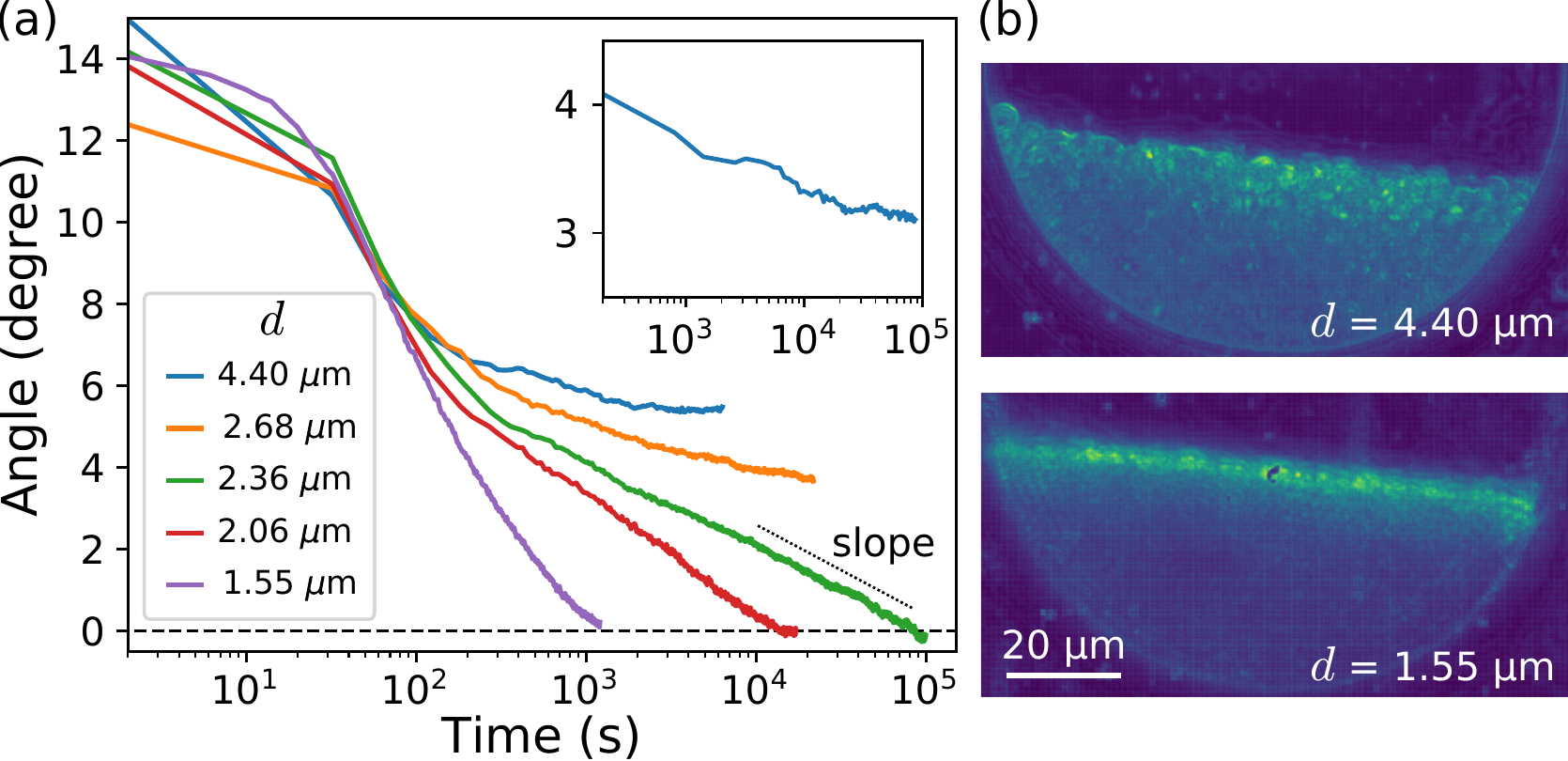}
\caption{\textbf{(a)} Long-time avalanches for different particle sizes, with initial angle $\theta_{start} = \SI{15}{\degree}$. The P\'{e}clet numbers corresponding to the different sizes are: $\mathrm{Pe}=397$ for \SI{4.40}{\micro\meter}, $\mathrm{Pe}=55$ for \SI{2.68}{\micro\meter}, $\mathrm{Pe}=33$ for \SI{2.36}{\micro\meter}, $\mathrm{Pe}=19$ for \SI{2.06}{\micro\meter}, and $\mathrm{Pe}=6$ for \SI{1.55}{\micro\meter}. \textit{Inset}: Long-time behavior for \SI{4.40}{\micro\meter} particles, with lower initial angle. \textbf{(b)} Heat map of particles displacement in the creep regime for two particle sizes. The map is obtained by computing image differences with a small time step (\SI{1}{\second} for \SI{1.55}{\micro\meter} particles and \SI{2}{\second} for \SI{4.40}{\micro\meter} particles) and averaging over \SI{1}{\minute}. The light colors correspond to area where the particles are moving. The creep flow is localised in a small layer of particles at the top of the pile (typically 1 or 2 particle sizes).\label{fig:avalanches_longtimes}}
\end{figure}

Finally, to study the effect of the P\'{e}clet number on the avalanche dynamics, we recorded several long-time avalanches with different particle sizes starting at the same initial inclination angle $\theta_{start}=\SI{15}{\degree}$. Note that very long measurements are not very easy to achieve because the samples usually dry in less than 48 hours. Data are presented on figure~\ref{fig:avalanches_longtimes}~(a).  The fast avalanche regimes seems rather independent of $\mathrm{Pe}$ as the different curves collapse pretty well on short times. On the contrary, the slow creep regime is dramatically slower when the P\'{e}clet is higher. Below $\theta_c$, we observe a relaxation of $\theta$ that is logarithmic in time on several decades.  When $\mathrm{Pe}$ is increased, the slope of this logarithmic regime clearly decreases, hence the time needed to reach $\theta = \SI{0}{\degree}$ becomes longer and longer. In the limit where $\mathrm{Pe} \gg 1$, we expect to retrieve the macroscopic granular behavior with no avalanche motion measured below the angle of repose. Finally, when the angle is very close to \SI{0}{\degree}, we observe a sharp transition to a stable regime with no average motion of the pile.\\
The region of the pile where particles are moving can be estimated by using bigger magnifications on the microscope (see figure~\ref{fig:avalanches_longtimes} (b)). Interestingly, the flow in the creep regime is limited to a small layer of grains (only one or two particle sizes) at the top of the pile.

\section{Creep regime model}

\begin{figure}[ht!]
\includegraphics[width=\columnwidth]{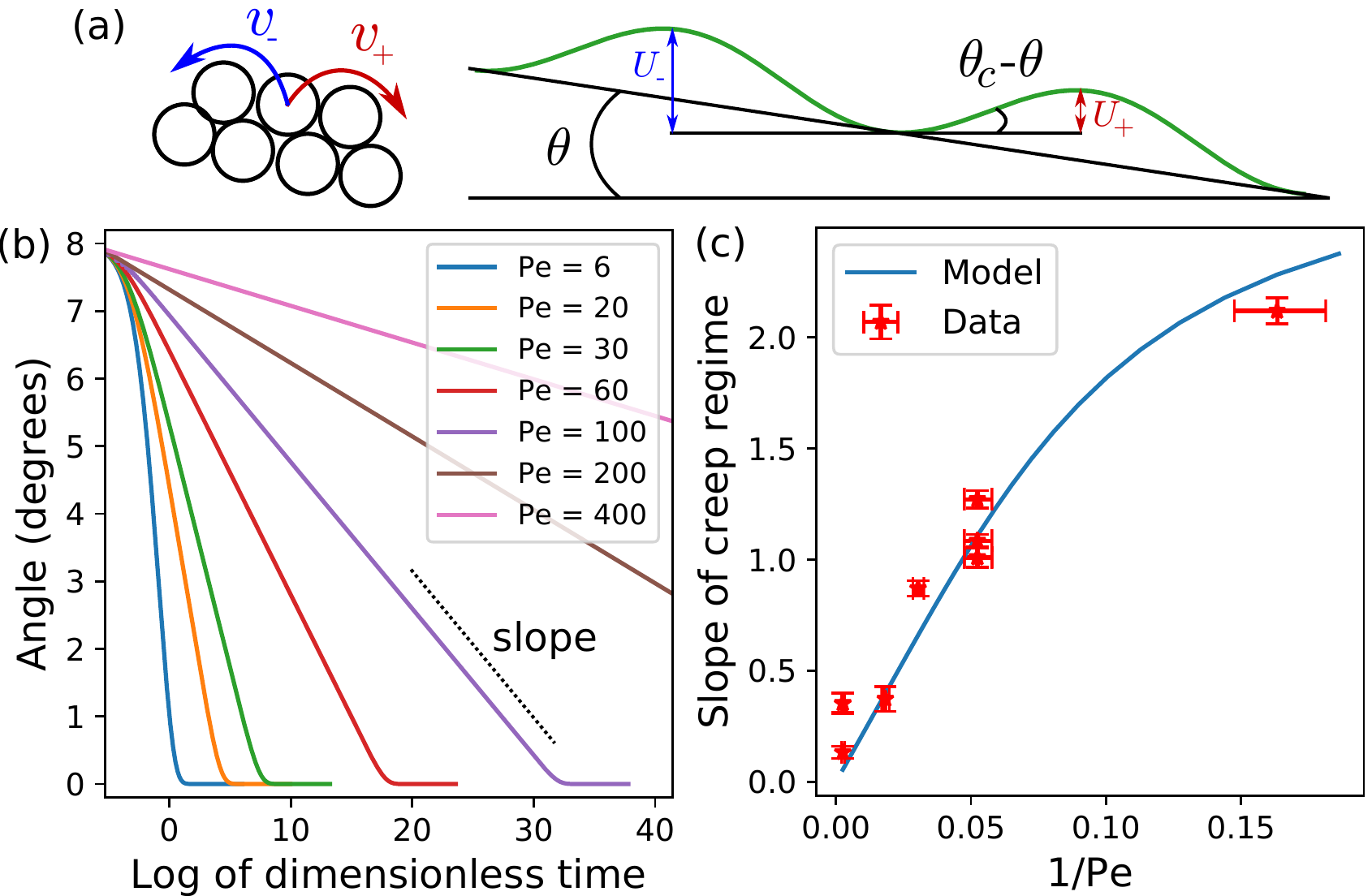}%
\caption{\textbf{(a)} Schematic representation of the potential felt by a particle at the top of the inclined pile. \textbf{(b)} Avalanches dynamics in the creep regimes obtained with the 1D model for different values of $\mathrm{Pe}$, as a function of $\log(t/\tau)$. The parameter $\alpha$ was taken equal to the best value that fits the experimental data: $\alpha = 2.64$. Each angle evolution can be characterized by the slope of its logarithmic regime. \textbf{(c)} Comparison of the logarithmic regime slopes between the experimental data and and the model with $\alpha = 2.64$. For the experimental data, each point corresponds to a sample. The horizontal errorbars are given by the dispersion of particle's diameters. The vertical errorbars are given by the dispersion of the slopes measured in the different drums of a each sample.\label{fig:model}}
\end{figure}

To retrieve these observations, we propose a very simple 1D model to describe the flow of particles in the creep regime. The model is based on the assumption that a bead at the top of the pile can be described as a Brownian particle in a rough energy landscape, that can cross barriers of potential energy $U$ with a probability $p$ that follows Kramer's theory: $p\propto \exp(-U/k_\mathrm{B}T)$~\cite{Kramers1940,Hanggi1990}. This situation is schematized on figure~\ref{fig:model} (a). Considering a particle going down the pile, we expect that the energy barrier it faces $U_+$ is proportional to the gravitational potential diminished by the tangent of the inclination angle $\theta$. It must also vanishes for $\theta = \theta_c$ because the particles are free to fall if the pile is inclined above the angle of repose. Similarly, the energy $U_-$ faced by a particle going up the pile is increased by $\tan \theta$. We then have:
\begin{equation}
U_\pm = \alpha mgd (\tan\theta_c \mp \tan\theta)
\end{equation}
where $\alpha$ is a dimensionless coefficient.\\
This leads to the following expressions for the average velocities of a particle going down the pile $v_+$ and of a particle going up the pile $v_-$:
\begin{equation}
v_\pm = f d \exp \left[ -\alpha \mathrm{Pe} (\tan\theta_c \mp \tan\theta) \right]
\end{equation}
where $f$ is a rate prefactor, which is dimensionally proportional to $\Delta \rho g d/\eta$, with $\eta$ the viscosity of the fluid.\\
By volume conservation during the flow, we have: $1/2 \times (D/2)^2 \mathrm{d}\theta/\mathrm{d}t = h(v_+ - v_-)$, where $h$ is the height of the layer of flowing particles. Assuming that $h \sim d$, we then obtain the law for the evolution of the pile angle:
\begin{equation}
\frac{\mathrm{d}\theta}{\mathrm{d}t} = -\frac{1}{\tau} \operatorname{e}^{-\alpha \mathrm{Pe} \tan \theta_c} \sinh \left(\alpha \mathrm{Pe} \tan \theta \right)
\end{equation}
where $\tau$ is a time scaling that depends on the particle's weight, the viscosity of the fluid, and the geometrical parameters of the drum: $\tau \propto D^2\eta/(\Delta\rho g d^3)$.\\
Since the angles remains small ($\theta \leq \theta_c \approx \SI{8}{\degree}$), we can use the approximation $\tan \theta \approx \theta$ and solve the equation with the initial condition $\theta (t=0) = \theta_c$:
\begin{equation}
\theta(t) = \frac{2}{\alpha\mathrm{Pe}} \, \mathrm{arcoth} \left( \operatorname{e}^{\frac{t}{\tau} \alpha \mathrm{Pe}} \,\coth (\alpha \mathrm{Pe} \theta_c /2) \, \operatorname{e}^{\operatorname{e}^{- \alpha\mathrm{Pe} \theta_c}}  \right)
\end{equation}
This solution is plotted as a function of $\log(t/\tau)$ for different values of $\mathrm{Pe}$ in figure \ref{fig:model} (b). The model retrieves the sharp change of regime when $\theta$ approaches \SI{0}{\degree}. It also retrieves the fact that the creep flow is logarithmic in time, with a slope that decrease when the P\'{e}clet is increased. In the limit of large $\mathrm{Pe}$, the solution can be simplified to:
\begin{equation}
\theta(t) \approx \theta_c - \frac{1}{\alpha\mathrm{Pe}} \log \left( \alpha \mathrm{Pe} \frac{t}{\tau} \right)
\end{equation}
as long as $\theta$ is not too close from $\theta_c$ nor \SI{0}{\degree}. As a consequence, in the limit of large P\'{e}clet numbers, the slope of the logarithmic regimes evolves as $1/\mathrm{Pe}$.

A comparison of the slopes measured in the logarithmic regimes of experimental avalanches and of the slopes extracted from the model is shown in figure~\ref{fig:model} (c). The value of $\alpha$ that gives the best agreement between the experiment and the model is $\alpha = 2.64$, but the agreement remains reasonable for $2 < \alpha < 3$. A value of $\alpha > 1$ can be interpreted as the fact that the barrier a particle needs to overcome is a bit higher than the diameter of its neighboring particle. This could be due to the fact that the pile itself is fluctuating, or to 3D effects not taken in account by our over-simplistic model.\\
The values of $\tau$ can be estimated by comparing the time needed for the pile to reach $\theta = \SI{0}{\degree}$ in the model and in the experiments. However, this quantity is difficult to extrapolate experimentally for slow avalanches, as a small error on the slope can lead to big errors on $\tau$. For the three smaller sizes (1.55, 2.06 and \SI{2.36}{\micro\meter}) we estimate that $\SI{600}{\second} \geq \tau \geq \SI{30}{\second}$. This order of magnitude is consistent with the scaling $\tau \propto D^2\eta/(\Delta\rho g d^3)$, that gives values between \SI{320}{\second} and \SI{90}{\second} for those bead sizes.

\section{Conclusion}

In conclusion, we have studied the avalanche dynamics of a ``Brownian granular material'', made of nearly frictionless grains small enough to be sensitive to thermal agitation. This granular materials exhibits a slow creep regime below its angle of repose, that does not exist in its athermal counterpart. In this regime, the pile angle shows a logarithmic relaxation to zero that dramatically depends on the P\'{e}clet number. This phenomenon is reminiscent of relaxations observed in macroscopic granular materials submitted to external perturbations, such as vibration~\cite{Knight1995,Sanchez2007}, thermal cycling~\cite{Divoux2008}, or air-injection below the fluidization threshold~\cite{Nguyen2011}. We propose a very simple 1D model to describe this regime. It gives good agreement with experimental data with only one free-parameter.\\
We believe that this kind of experimental system could bring useful insights in the field of dense Brownian suspensions rheology. For example, an athermal analogue of these suspensions, with an effective logarithmic repulsive potential representing the vibrational entropic forces was recently introduced~\cite{Trulsson2015}. However, it remains unclear that this kind of approach could take account of the experimental measurements reported here, since the thermal agitation is crucial to observe the creep regime. Our study provides a first step to make a link between the avalanche dynamics and the rheology of Brownian granular materials. Future works could focus on flow geometries that enable to control the applied stress, such as the inclined plane configuration widely used in granular flows.

\section{Acknowledgments}

\begin{acknowledgments}
This work was supported by the European Research Council (ERC) under the European Union Horizon 2020 Research and Innovation program (ERC Grant 647384) and by the Labex MEC (ANR-10-LABX-0092) under the A*MIDEX project (ANR-11-IDEX-0001-02) funded by the French government program Investissements d'Avenir. The authors want to thank Igor Ozerov for his help during
the fabrication of the microfluidic system. Data presented in this article are available in Zenodo repository: \href{https://doi.org/10.5281/zenodo.1035785}{10.5281/zenodo.1035785}.
\end{acknowledgments}

\bibliography{biblio}

\begin{thebibliography}{24}%
\makeatletter
\providecommand \@ifxundefined [1]{%
 \@ifx{#1\undefined}
}%
\providecommand \@ifnum [1]{%
 \ifnum #1\expandafter \@firstoftwo
 \else \expandafter \@secondoftwo
 \fi
}%
\providecommand \@ifx [1]{%
 \ifx #1\expandafter \@firstoftwo
 \else \expandafter \@secondoftwo
 \fi
}%
\providecommand \natexlab [1]{#1}%
\providecommand \enquote  [1]{``#1''}%
\providecommand \bibnamefont  [1]{#1}%
\providecommand \bibfnamefont [1]{#1}%
\providecommand \citenamefont [1]{#1}%
\providecommand \href@noop [0]{\@secondoftwo}%
\providecommand \href [0]{\begingroup \@sanitize@url \@href}%
\providecommand \@href[1]{\@@startlink{#1}\@@href}%
\providecommand \@@href[1]{\endgroup#1\@@endlink}%
\providecommand \@sanitize@url [0]{\catcode `\\12\catcode `\$12\catcode
  `\&12\catcode `\#12\catcode `\^12\catcode `\_12\catcode `\%12\relax}%
\providecommand \@@startlink[1]{}%
\providecommand \@@endlink[0]{}%
\providecommand \url  [0]{\begingroup\@sanitize@url \@url }%
\providecommand \@url [1]{\endgroup\@href {#1}{\urlprefix }}%
\providecommand \urlprefix  [0]{URL }%
\providecommand \Eprint [0]{\href }%
\providecommand \doibase [0]{http://dx.doi.org/}%
\providecommand \selectlanguage [0]{\@gobble}%
\providecommand \bibinfo  [0]{\@secondoftwo}%
\providecommand \bibfield  [0]{\@secondoftwo}%
\providecommand \translation [1]{[#1]}%
\providecommand \BibitemOpen [0]{}%
\providecommand \bibitemStop [0]{}%
\providecommand \bibitemNoStop [0]{.\EOS\space}%
\providecommand \EOS [0]{\spacefactor3000\relax}%
\providecommand \BibitemShut  [1]{\csname bibitem#1\endcsname}%
\let\auto@bib@innerbib\@empty
\bibitem [{\citenamefont {Courrech~du Pont}\ \emph {et~al.}(2003)\citenamefont
  {Courrech~du Pont}, \citenamefont {Gondret}, \citenamefont {Perrin},\ and\
  \citenamefont {Rabaud}}]{CourrechduPont2003}%
  \BibitemOpen
  \bibfield  {author} {\bibinfo {author} {\bibfnamefont {S.}~\bibnamefont
  {Courrech~du Pont}}, \bibinfo {author} {\bibfnamefont {P.}~\bibnamefont
  {Gondret}}, \bibinfo {author} {\bibfnamefont {B.}~\bibnamefont {Perrin}}, \
  and\ \bibinfo {author} {\bibfnamefont {M.}~\bibnamefont {Rabaud}},\ }\href
  {\doibase 10.1103/PhysRevLett.90.044301} {\bibfield  {journal} {\bibinfo
  {journal} {Phys. Rev. Lett.}\ }\textbf {\bibinfo {volume} {90}},\ \bibinfo
  {pages} {044301} (\bibinfo {year} {2003})}\BibitemShut {NoStop}%
\bibitem [{\citenamefont {Pailha}\ \emph {et~al.}(2008)\citenamefont {Pailha},
  \citenamefont {Nicolas},\ and\ \citenamefont {Pouliquen}}]{Pailha2008}%
  \BibitemOpen
  \bibfield  {author} {\bibinfo {author} {\bibfnamefont {M.}~\bibnamefont
  {Pailha}}, \bibinfo {author} {\bibfnamefont {M.}~\bibnamefont {Nicolas}}, \
  and\ \bibinfo {author} {\bibfnamefont {O.}~\bibnamefont {Pouliquen}},\ }\href
  {\doibase 10.1063/1.3013896} {\bibfield  {journal} {\bibinfo  {journal}
  {Physics of Fluids}\ }\textbf {\bibinfo {volume} {20}},\ \bibinfo {pages}
  {111701} (\bibinfo {year} {2008})}\BibitemShut {NoStop}%
\bibitem [{\citenamefont {Rondon}\ \emph {et~al.}(2011)\citenamefont {Rondon},
  \citenamefont {Pouliquen},\ and\ \citenamefont {Aussillous}}]{Rondon2011}%
  \BibitemOpen
  \bibfield  {author} {\bibinfo {author} {\bibfnamefont {L.}~\bibnamefont
  {Rondon}}, \bibinfo {author} {\bibfnamefont {O.}~\bibnamefont {Pouliquen}}, \
  and\ \bibinfo {author} {\bibfnamefont {P.}~\bibnamefont {Aussillous}},\
  }\href {\doibase 10.1063/1.3594200} {\bibfield  {journal} {\bibinfo
  {journal} {Physics of Fluids}\ }\textbf {\bibinfo {volume} {23}},\ \bibinfo
  {pages} {073301} (\bibinfo {year} {2011})}\BibitemShut {NoStop}%
\bibitem [{\citenamefont {Piazza}(2014)}]{Piazza2014}%
  \BibitemOpen
  \bibfield  {author} {\bibinfo {author} {\bibfnamefont {R.}~\bibnamefont
  {Piazza}},\ }\href {http://stacks.iop.org/0034-4885/77/i=5/a=056602}
  {\bibfield  {journal} {\bibinfo  {journal} {Reports on Progress in Physics}\
  }\textbf {\bibinfo {volume} {77}},\ \bibinfo {pages} {056602} (\bibinfo
  {year} {2014})}\BibitemShut {NoStop}%
\bibitem [{\citenamefont {Royall}\ \emph {et~al.}(2005)\citenamefont {Royall},
  \citenamefont {van Roij},\ and\ \citenamefont {van Blaaderen}}]{Royall2005}%
  \BibitemOpen
  \bibfield  {author} {\bibinfo {author} {\bibfnamefont {C.~P.}\ \bibnamefont
  {Royall}}, \bibinfo {author} {\bibfnamefont {R.}~\bibnamefont {van Roij}}, \
  and\ \bibinfo {author} {\bibfnamefont {A.}~\bibnamefont {van Blaaderen}},\
  }\href {http://stacks.iop.org/0953-8984/17/i=15/a=005} {\bibfield  {journal}
  {\bibinfo  {journal} {Journal of Physics: Condensed Matter}\ }\textbf
  {\bibinfo {volume} {17}},\ \bibinfo {pages} {2315} (\bibinfo {year}
  {2005})}\BibitemShut {NoStop}%
\bibitem [{\citenamefont {Dullens}\ \emph {et~al.}(2006)\citenamefont
  {Dullens}, \citenamefont {Aarts},\ and\ \citenamefont {Kegel}}]{Dullens2006}%
  \BibitemOpen
  \bibfield  {author} {\bibinfo {author} {\bibfnamefont {R.~P.~A.}\
  \bibnamefont {Dullens}}, \bibinfo {author} {\bibfnamefont {D.~G. A.~L.}\
  \bibnamefont {Aarts}}, \ and\ \bibinfo {author} {\bibfnamefont {W.~K.}\
  \bibnamefont {Kegel}},\ }\href {\doibase 10.1103/PhysRevLett.97.228301}
  {\bibfield  {journal} {\bibinfo  {journal} {Phys. Rev. Lett.}\ }\textbf
  {\bibinfo {volume} {97}},\ \bibinfo {pages} {228301} (\bibinfo {year}
  {2006})}\BibitemShut {NoStop}%
\bibitem [{\citenamefont {Thorneywork}\ \emph {et~al.}(2017)\citenamefont
  {Thorneywork}, \citenamefont {Abbott}, \citenamefont {Aarts},\ and\
  \citenamefont {Dullens}}]{Thorneywork2017}%
  \BibitemOpen
  \bibfield  {author} {\bibinfo {author} {\bibfnamefont {A.~L.}\ \bibnamefont
  {Thorneywork}}, \bibinfo {author} {\bibfnamefont {J.~L.}\ \bibnamefont
  {Abbott}}, \bibinfo {author} {\bibfnamefont {D.~G. A.~L.}\ \bibnamefont
  {Aarts}}, \ and\ \bibinfo {author} {\bibfnamefont {R.~P.~A.}\ \bibnamefont
  {Dullens}},\ }\href {\doibase 10.1103/PhysRevLett.118.158001} {\bibfield
  {journal} {\bibinfo  {journal} {Phys. Rev. Lett.}\ }\textbf {\bibinfo
  {volume} {118}},\ \bibinfo {pages} {158001} (\bibinfo {year}
  {2017})}\BibitemShut {NoStop}%
\bibitem [{\citenamefont {Russel}\ \emph {et~al.}(1989)\citenamefont {Russel},
  \citenamefont {Saville},\ and\ \citenamefont {Schowalter}}]{Russel1989}%
  \BibitemOpen
  \bibfield  {author} {\bibinfo {author} {\bibfnamefont {W.~B.}\ \bibnamefont
  {Russel}}, \bibinfo {author} {\bibfnamefont {D.~A.}\ \bibnamefont {Saville}},
  \ and\ \bibinfo {author} {\bibfnamefont {W.~R.}\ \bibnamefont {Schowalter}},\
  }\href@noop {} {\emph {\bibinfo {title} {Colloidal dispersions}}}\ (\bibinfo
  {publisher} {Cambridge university press},\ \bibinfo {year}
  {1989})\BibitemShut {NoStop}%
\bibitem [{\citenamefont {Mewis}\ and\ \citenamefont
  {Wagner}(2012)}]{Mewis2012}%
  \BibitemOpen
  \bibfield  {author} {\bibinfo {author} {\bibfnamefont {J.}~\bibnamefont
  {Mewis}}\ and\ \bibinfo {author} {\bibfnamefont {N.~J.}\ \bibnamefont
  {Wagner}},\ }\href@noop {} {\emph {\bibinfo {title} {Colloidal suspension
  rheology}}}\ (\bibinfo  {publisher} {Cambridge University Press},\ \bibinfo
  {year} {2012})\BibitemShut {NoStop}%
\bibitem [{\citenamefont {{Gayvallet, H.}}\ and\ \citenamefont {{G\'{e}minard,
  J.-C.}}(2002)}]{Gayvallet2002}%
  \BibitemOpen
  \bibfield  {author} {\bibinfo {author} {\bibnamefont {{Gayvallet, H.}}}\ and\
  \bibinfo {author} {\bibnamefont {{G\'{e}minard, J.-C.}}},\ }\href {\doibase
  10.1140/epjb/e2002-00391-6} {\bibfield  {journal} {\bibinfo  {journal} {Eur.
  Phys. J. B}\ }\textbf {\bibinfo {volume} {30}},\ \bibinfo {pages} {369}
  (\bibinfo {year} {2002})}\BibitemShut {NoStop}%
\bibitem [{\citenamefont {Duffy}\ \emph {et~al.}(1998)\citenamefont {Duffy},
  \citenamefont {McDonald}, \citenamefont {Schueller},\ and\ \citenamefont
  {Whitesides}}]{Duffy1998}%
  \BibitemOpen
  \bibfield  {author} {\bibinfo {author} {\bibfnamefont {D.~C.}\ \bibnamefont
  {Duffy}}, \bibinfo {author} {\bibfnamefont {J.~C.}\ \bibnamefont {McDonald}},
  \bibinfo {author} {\bibfnamefont {O.~J.~A.}\ \bibnamefont {Schueller}}, \
  and\ \bibinfo {author} {\bibfnamefont {G.~M.}\ \bibnamefont {Whitesides}},\
  }\href {\doibase 10.1021/ac980656z} {\bibfield  {journal} {\bibinfo
  {journal} {Analytical Chemistry}\ }\textbf {\bibinfo {volume} {70}},\
  \bibinfo {pages} {4974} (\bibinfo {year} {1998})},\ \bibinfo {note} {pMID:
  21644679}\BibitemShut {NoStop}%
\bibitem [{\citenamefont {McDonald}\ and\ \citenamefont
  {Whitesides}(2002)}]{McDonald2002}%
  \BibitemOpen
  \bibfield  {author} {\bibinfo {author} {\bibfnamefont {J.~C.}\ \bibnamefont
  {McDonald}}\ and\ \bibinfo {author} {\bibfnamefont {G.~M.}\ \bibnamefont
  {Whitesides}},\ }\href {\doibase 10.1021/ar010110q} {\bibfield  {journal}
  {\bibinfo  {journal} {Accounts of Chemical Research}\ }\textbf {\bibinfo
  {volume} {35}},\ \bibinfo {pages} {491} (\bibinfo {year} {2002})},\ \bibinfo
  {note} {pMID: 12118988}\BibitemShut {NoStop}%
\bibitem [{\citenamefont {Friend}\ and\ \citenamefont
  {Yeo}(2010)}]{Friend2010}%
  \BibitemOpen
  \bibfield  {author} {\bibinfo {author} {\bibfnamefont {J.}~\bibnamefont
  {Friend}}\ and\ \bibinfo {author} {\bibfnamefont {L.}~\bibnamefont {Yeo}},\
  }\href {\doibase 10.1063/1.3259624} {\bibfield  {journal} {\bibinfo
  {journal} {Biomicrofluidics}\ }\textbf {\bibinfo {volume} {4}},\ \bibinfo
  {pages} {026502} (\bibinfo {year} {2010})}\BibitemShut {NoStop}%
\bibitem [{\citenamefont {Andreotti}\ \emph {et~al.}(2013)\citenamefont
  {Andreotti}, \citenamefont {Forterre},\ and\ \citenamefont
  {Pouliquen}}]{Andreotti2013}%
  \BibitemOpen
  \bibfield  {author} {\bibinfo {author} {\bibfnamefont {B.}~\bibnamefont
  {Andreotti}}, \bibinfo {author} {\bibfnamefont {Y.}~\bibnamefont {Forterre}},
  \ and\ \bibinfo {author} {\bibfnamefont {O.}~\bibnamefont {Pouliquen}},\
  }\href@noop {} {\emph {\bibinfo {title} {Granular media: between fluid and
  solid}}}\ (\bibinfo  {publisher} {Cambridge University Press},\ \bibinfo
  {year} {2013})\BibitemShut {NoStop}%
\bibitem [{\citenamefont {Peyneau}\ and\ \citenamefont
  {Roux}(2008)}]{Peyneau2008}%
  \BibitemOpen
  \bibfield  {author} {\bibinfo {author} {\bibfnamefont {P.-E.}\ \bibnamefont
  {Peyneau}}\ and\ \bibinfo {author} {\bibfnamefont {J.-N.}\ \bibnamefont
  {Roux}},\ }\href {\doibase 10.1103/PhysRevE.78.011307} {\bibfield  {journal}
  {\bibinfo  {journal} {Phys. Rev. E}\ }\textbf {\bibinfo {volume} {78}},\
  \bibinfo {pages} {011307} (\bibinfo {year} {2008})}\BibitemShut {NoStop}%
\bibitem [{\citenamefont {Israelachvili}(2011)}]{Israelachvili2011}%
  \BibitemOpen
  \bibfield  {author} {\bibinfo {author} {\bibfnamefont {J.~N.}\ \bibnamefont
  {Israelachvili}},\ }in\ \href {\doibase
  https://doi.org/10.1016/B978-0-12-375182-9.10014-4} {\emph {\bibinfo
  {booktitle} {Intermolecular and Surface Forces (Third Edition)}}},\ \bibinfo
  {editor} {edited by\ \bibinfo {editor} {\bibfnamefont {J.~N.}\ \bibnamefont
  {Israelachvili}}}\ (\bibinfo  {publisher} {Academic Press},\ \bibinfo
  {address} {San Diego},\ \bibinfo {year} {2011})\ \bibinfo {edition} {third
  edition}\ ed.,\ pp.\ \bibinfo {pages} {291 -- 340}\BibitemShut {NoStop}%
\bibitem [{\citenamefont {Clavaud}\ \emph {et~al.}(2017)\citenamefont
  {Clavaud}, \citenamefont {B\'{e}rut}, \citenamefont {Metzger},\ and\
  \citenamefont {Forterre}}]{Clavaud2017}%
  \BibitemOpen
  \bibfield  {author} {\bibinfo {author} {\bibfnamefont {C.}~\bibnamefont
  {Clavaud}}, \bibinfo {author} {\bibfnamefont {A.}~\bibnamefont {B\'{e}rut}},
  \bibinfo {author} {\bibfnamefont {B.}~\bibnamefont {Metzger}}, \ and\
  \bibinfo {author} {\bibfnamefont {Y.}~\bibnamefont {Forterre}},\ }\href
  {\doibase 10.1073/pnas.1703926114} {\bibfield  {journal} {\bibinfo  {journal}
  {Proceedings of the National Academy of Sciences}\ }\textbf {\bibinfo
  {volume} {114}},\ \bibinfo {pages} {5147} (\bibinfo {year}
  {2017})}\BibitemShut {NoStop}%
\bibitem [{\citenamefont {Kramers}(1940)}]{Kramers1940}%
  \BibitemOpen
  \bibfield  {author} {\bibinfo {author} {\bibfnamefont {H.}~\bibnamefont
  {Kramers}},\ }\href {\doibase https://doi.org/10.1016/S0031-8914(40)90098-2}
  {\bibfield  {journal} {\bibinfo  {journal} {Physica}\ }\textbf {\bibinfo
  {volume} {7}},\ \bibinfo {pages} {284 } (\bibinfo {year} {1940})}\BibitemShut
  {NoStop}%
\bibitem [{\citenamefont {H\"anggi}\ \emph {et~al.}(1990)\citenamefont
  {H\"anggi}, \citenamefont {Talkner},\ and\ \citenamefont
  {Borkovec}}]{Hanggi1990}%
  \BibitemOpen
  \bibfield  {author} {\bibinfo {author} {\bibfnamefont {P.}~\bibnamefont
  {H\"anggi}}, \bibinfo {author} {\bibfnamefont {P.}~\bibnamefont {Talkner}}, \
  and\ \bibinfo {author} {\bibfnamefont {M.}~\bibnamefont {Borkovec}},\ }\href
  {\doibase 10.1103/RevModPhys.62.251} {\bibfield  {journal} {\bibinfo
  {journal} {Rev. Mod. Phys.}\ }\textbf {\bibinfo {volume} {62}},\ \bibinfo
  {pages} {251} (\bibinfo {year} {1990})}\BibitemShut {NoStop}%
\bibitem [{\citenamefont {Knight}\ \emph {et~al.}(1995)\citenamefont {Knight},
  \citenamefont {Fandrich}, \citenamefont {Lau}, \citenamefont {Jaeger},\ and\
  \citenamefont {Nagel}}]{Knight1995}%
  \BibitemOpen
  \bibfield  {author} {\bibinfo {author} {\bibfnamefont {J.~B.}\ \bibnamefont
  {Knight}}, \bibinfo {author} {\bibfnamefont {C.~G.}\ \bibnamefont
  {Fandrich}}, \bibinfo {author} {\bibfnamefont {C.~N.}\ \bibnamefont {Lau}},
  \bibinfo {author} {\bibfnamefont {H.~M.}\ \bibnamefont {Jaeger}}, \ and\
  \bibinfo {author} {\bibfnamefont {S.~R.}\ \bibnamefont {Nagel}},\ }\href
  {\doibase 10.1103/PhysRevE.51.3957} {\bibfield  {journal} {\bibinfo
  {journal} {Phys. Rev. E}\ }\textbf {\bibinfo {volume} {51}},\ \bibinfo
  {pages} {3957} (\bibinfo {year} {1995})}\BibitemShut {NoStop}%
\bibitem [{\citenamefont {S\'anchez}\ \emph {et~al.}(2007)\citenamefont
  {S\'anchez}, \citenamefont {Raynaud}, \citenamefont {Lanuza}, \citenamefont
  {Andreotti}, \citenamefont {Cl\'ement},\ and\ \citenamefont
  {Aranson}}]{Sanchez2007}%
  \BibitemOpen
  \bibfield  {author} {\bibinfo {author} {\bibfnamefont {I.}~\bibnamefont
  {S\'anchez}}, \bibinfo {author} {\bibfnamefont {F.}~\bibnamefont {Raynaud}},
  \bibinfo {author} {\bibfnamefont {J.}~\bibnamefont {Lanuza}}, \bibinfo
  {author} {\bibfnamefont {B.}~\bibnamefont {Andreotti}}, \bibinfo {author}
  {\bibfnamefont {E.}~\bibnamefont {Cl\'ement}}, \ and\ \bibinfo {author}
  {\bibfnamefont {I.~S.}\ \bibnamefont {Aranson}},\ }\href {\doibase
  10.1103/PhysRevE.76.060301} {\bibfield  {journal} {\bibinfo  {journal} {Phys.
  Rev. E}\ }\textbf {\bibinfo {volume} {76}},\ \bibinfo {pages} {060301}
  (\bibinfo {year} {2007})}\BibitemShut {NoStop}%
\bibitem [{\citenamefont {Divoux}\ \emph {et~al.}(2008)\citenamefont {Divoux},
  \citenamefont {Gayvallet},\ and\ \citenamefont {G\'eminard}}]{Divoux2008}%
  \BibitemOpen
  \bibfield  {author} {\bibinfo {author} {\bibfnamefont {T.}~\bibnamefont
  {Divoux}}, \bibinfo {author} {\bibfnamefont {H.}~\bibnamefont {Gayvallet}}, \
  and\ \bibinfo {author} {\bibfnamefont {J.-C.}\ \bibnamefont {G\'eminard}},\
  }\href {\doibase 10.1103/PhysRevLett.101.148303} {\bibfield  {journal}
  {\bibinfo  {journal} {Phys. Rev. Lett.}\ }\textbf {\bibinfo {volume} {101}},\
  \bibinfo {pages} {148303} (\bibinfo {year} {2008})}\BibitemShut {NoStop}%
\bibitem [{\citenamefont {Nguyen}\ \emph {et~al.}(2011)\citenamefont {Nguyen},
  \citenamefont {Darnige}, \citenamefont {Bruand},\ and\ \citenamefont
  {Clement}}]{Nguyen2011}%
  \BibitemOpen
  \bibfield  {author} {\bibinfo {author} {\bibfnamefont {V.~B.}\ \bibnamefont
  {Nguyen}}, \bibinfo {author} {\bibfnamefont {T.}~\bibnamefont {Darnige}},
  \bibinfo {author} {\bibfnamefont {A.}~\bibnamefont {Bruand}}, \ and\ \bibinfo
  {author} {\bibfnamefont {E.}~\bibnamefont {Clement}},\ }\href {\doibase
  10.1103/PhysRevLett.107.138303} {\bibfield  {journal} {\bibinfo  {journal}
  {Phys. Rev. Lett.}\ }\textbf {\bibinfo {volume} {107}},\ \bibinfo {pages}
  {138303} (\bibinfo {year} {2011})}\BibitemShut {NoStop}%
\bibitem [{\citenamefont {Trulsson}\ \emph {et~al.}(2015)\citenamefont
  {Trulsson}, \citenamefont {Bouzid}, \citenamefont {Kurchan}, \citenamefont
  {Cl\'{e}ment}, \citenamefont {Claudin},\ and\ \citenamefont
  {Andreotti}}]{Trulsson2015}%
  \BibitemOpen
  \bibfield  {author} {\bibinfo {author} {\bibfnamefont {M.}~\bibnamefont
  {Trulsson}}, \bibinfo {author} {\bibfnamefont {M.}~\bibnamefont {Bouzid}},
  \bibinfo {author} {\bibfnamefont {J.}~\bibnamefont {Kurchan}}, \bibinfo
  {author} {\bibfnamefont {E.}~\bibnamefont {Cl\'{e}ment}}, \bibinfo {author}
  {\bibfnamefont {P.}~\bibnamefont {Claudin}}, \ and\ \bibinfo {author}
  {\bibfnamefont {B.}~\bibnamefont {Andreotti}},\ }\href
  {http://stacks.iop.org/0295-5075/111/i=1/a=18001} {\bibfield  {journal}
  {\bibinfo  {journal} {EPL (Europhysics Letters)}\ }\textbf {\bibinfo {volume}
  {111}},\ \bibinfo {pages} {18001} (\bibinfo {year} {2015})}\BibitemShut
  {NoStop}%
\end{thebibliography}%

\end{document}